\documentstyle{mn}
\input{psfig}
\def\eta{et al. }
\def\lx{$L_{\rm x}$ }
\def\lt{$L_{\rm r,total}$ }
\def\lc{$L_{\rm r,core}$ }
\def\ergs{erg cm$^{-2}$ s$^{-1}$ }

\title[Soft X-ray properties of radio sources]{The soft X-ray properties
of a complete sample of radio sources}

\author[J. Siebert \eta]{J. Siebert,$^1$ W. Brinkmann,$^1$ R. Morganti,$^{2,3}$
C.N. Tadhunter,$^4$ I.J. Danziger,$^5$ \cr
R.A.E. Fosbury$^{6,7}$ and S. di Serego Alighieri$^8$\\
$^1$ Max-Planck-Institut f\"ur Extraterrestrische Physik,
Giessenbachstrasse, D-85740 Garching, Germany\\
$^2$ Istituto di Radioastronomia, CNR,
via Gobetti 101, I-40129 Bologna, Italy\\
$^3$  Australia Telescope National
Facility, CSIRO, PO Box 76, Epping, NSW 2121, Australia\\
$^4$  Department of Physics, University of Sheffield, Sheffield S3\\
$^5$ ESO, Karl Schwarzschild Strasse 2, D-85748 Garching, Germany\\
$^6$  Space Telescope-European Coordinating Facility, Karl
Schwarzschild Strasse 2, D-85748 Garching, Germany\\
$^7$ Affiliated to the Astrophysics Division, Space Science Department,
European Space Agency\\
$^8$  Osservatorio Astrofisico
di Arcetri, Largo E. Fermi 5, 50125 Firenze, Italy
 }
\date{Accepted 1995 November 24. Received 1995 November 9; in original form
1995 March 7}
\pubyear{1995}

\begin{document}

\maketitle
 
\begin{abstract}
 
We present the soft X-ray (0.1--2.4 keV) properties of a complete sample of 88
southern radio sources derived from the Wall \& Peacock 2-Jy sample. It 
comprises 68 radio galaxies, 18 quasars and 2 BL Lac objects. Whereas both BL 
Lac objects and all but one quasar are detected in the {\it ROSAT} All-Sky 
Survey, the fraction of detected radio galaxies is only $\sim$ 60 per cent. 
For the undetected sources upper limits to the X-ray flux are given. 
We confirm the correlation of the soft X-ray luminosity (\lx) with the core 
radio luminosity (\lc) for galaxies as well as for quasars using partial 
correlation analysis, whereas the corresponding correlations between \lx and
\lt are probably spurious due to sample selection effects. We also find 
strong correlations between \lx and \lc for both Fanaroff-Riley type I 
(FR~I) and type II galaxies. The broad-line radio galaxies (BLRGs) and the 
quasars are at the top end of the X-ray luminosity distribution and the 
detection rate of these objects generally is higher than that of the narrow- 
or weak-lined radio galaxies. This indicates the presence of an anisotropic 
X-ray component in BLRGs and quasars, as predicted by unified schemes for 
radio sources. 

\end{abstract}

 
\section{Introduction}

There is growing evidence that much of the apparent diversity  
found amongst active galactic nuclei (AGNs) can be explained by 
anisotropic obscuration and, in the case of radio-loud objects, by 
relativistic boosting effects resulting from a high-velocity jet. Within 
this unifying framework, the classification of an AGN is believed to 
be influenced strongly by the orientation with respect to the line of 
sight. The subject has been discussed extensively in recent years, 
for both radio-quiet and radio-loud AGNs (see Antonucci 1993 for a 
review). Well-defined samples of AGNs with comprehensive observational data 
on the relevant parameters at different wavelengths are essential 
for statistically reliable tests of these theories.

In the X-ray waveband, statistical properties of both radio galaxies 
and quasars have been studied using data obtained mostly with the 
Imaging Proportional Counter (IPC) on board the {\it Einstein} satellite. 
Feigelson \& Berg (1983), using a heterogeneous sample of radio 
galaxies, found a correlation between X-ray luminosity and both total 
and core radio power. The optical emission lines do not, however, 
appear to be a good predictor of the X-ray emission. According to 
these authors, the results can be best explained by assuming that the X-ray 
emission comes from the hot gas of a putative surrounding cluster. For a 
sample of forty 3C radio galaxies, Fabbiano et al. (1984) found that FR~II 
galaxies and  objects with strong optical emission-line spectra tend to be 
more powerful in X-rays than those with an FR~I morphology and weak or absent 
optical emission lines. In addition, they confirm a strong correlation 
between the radio core power and X-ray luminosity. They 
conclude, in contradiction to Feigelson \& Berg (1983), that the X-ray 
emission in radio galaxies is dominated by the AGN rather than diffuse hot gas.

The key issue at hand, therefore, concerning our understanding of the 
X-ray/radio/optical correlations is that of distinguishing between 
these alternative origins for the X-ray emission.

In the case of quasars, early studies (see e.g., Zamorani et al. 1984; Wilkes 
\& Elvis 1987) showed that two components contribute to the X-ray emission
of radio-loud objects. One is isotropic and likely to 
be present also in radio-quiet quasars. The other appears to be 
associated with the radio core emission (e.g., by the synchrotron 
self-Compton mechanism) and is probably beamed. Browne \& Murphy 
(1987) and Kembhavi, Feigelson \& Singh (1986) have found strong correlations 
between the X-ray luminosity and both the extended and the core radio power. 

The slope of the correlation between X-ray luminosity and the extended 
radio power becomes flatter for the less core-dominated radio sources, 
i.e., for a given extended radio luminosity, the core-dominated 
sources have significantly stronger X-ray emission. Browne \& Murphy (1987)
have shown that this behaviour supports the idea of two X-ray 
components with the beamed one being prominent in the core-dominated 
quasars. As further confirmation of this, more recent studies of the 
spectral slope have shown that in radio-loud quasars the soft X-ray 
spectrum is correlated with the core dominance parameter $R$: the 
spectrum gets flatter as $R$ increases (Shastri \eta 1993; Wilkes 1994). 

However, none of the studies includes an analysis of a complete 
sample of both radio galaxies and quasars for which good optical 
spectroscopic data as well as X-ray and radio 
observations are available. Recently, a complete sample with such 
characteristics has been established: it includes 88 southern radio sources 
and represents a subsample of the Wall \& Peacock 2.7-GHz catalogue (Wall \& 
Peacock 1985; Morganti, Killeen \& Tadhunter 1993; Tadhunter \eta 1993; di 
Serego Alighieri \eta 1994). Here we present the soft X-ray properties of 
these sources using data from both the {\it ROSAT} All-Sky Survey (Voges 1993) 
and pointed {\it ROSAT} observations. Throughout this paper we assume $H_{\rm 0} 
=50$ km s$^{-1}$ Mpc$^{-1}$ and $q_{\rm 0} = 0$. 

\section{The Sample}

Our subsample is defined by redshift {\it z} $<$ 0.7, declination $\delta < 
10^{\rm o}$, and it is complete down to a flux density level of 2 Jy at 2.7 
GHz. Optical spectra of all sources, including those of the original Wall \& 
Peacock sample with $\delta<10^{\circ}$ and no spectroscopic redshift, 
were taken by Tadhunter \eta (1993) and di Serego Alighieri \eta (1994)
with the ESO 3.6-m and 2.2-m telescopes at La Silla, Chile. The redshift 
constraint was chosen to obtain accurate measurements of the 
[O\,{\sevensize III}]$\lambda$5007 line flux for all objects. In addition, 
the [O\,{\sevensize II}]$\lambda$3727 and H$\beta$ emission-line fluxes were 
determined to make use of ionization diagnostics. For observational details 
and the presentation of the results, see Tadhunter \eta (1993) and di Serego 
Alighieri \eta (1994). 

Radio data from the Very Large Array (VLA) and the Australia Telescope 
Compact Array (ATCA) have been obtained at 5~GHz with an angular 
resolution of $\sim 3$ arcsec (Morganti et al. 1993). 
These data and information from literature (Stickel, Meisenheimer \& K\"uhr 
1994; Zirbel \& Baum 1995) have allowed a morphological classification of the 
sources according to the Fanaroff \& Riley (1974) convention. In some cases 
the FR-type could not unambiguously be determined,
which is due to either inadequate radio data or the presence of transition 
sources which show morphological details typical of both FR~I and FR~II (e.g., 
PKS 1333$-$33 or Her A). The adopted classification for this study is given in 
Table \ref{data}. For most of the sources core flux densities are available 
from literature (see references given in the notes to Table \ref{data}). 

\section{Data analysis}
\subsection{ROSAT All-Sky Survey}

\begin{table*}
 \begin{minipage}{15.5cm}
 \vbox to220mm{\vfil Landscape table to go here.
 \caption{\label{data}} \vfil}
 \begin{tabular}{l}
 \end{tabular}
 \end{minipage}
\end{table*}

We have examined the {\it ROSAT} All-Sky Survey (RASS) data for the sources 
in the sample in order to determine their soft X-ray properties or to derive 
useful upper limit fluxes in the case of non-detections. This has 
been done using a procedure based on standard commands within the 
{\sevensize EXSAS} environment (Zimmermann et al. 1993). The procedure 
uses a maximum-likelihood source detection algorithm which returns the 
likelihood of existence for a X-ray source at the specified radio position, 
the number of source photons within 5 times the FWHM of the Position 
Sensitive Proportional Counter (PSPC) point spread function (PSF) and the 
error in the number of source photons. For the RASS the FWHM of the PSPC 
point spread function is estimated to be $\sim 60$ arcsec (Zimmermann et al. 
1993)

We consider a radio source to be detected in soft X-rays if the likelihood of 
existence is greater than 5.91, which corresponds to $3\sigma$. Thus, the 
statistical probability of identifying a background fluctuation as a source 
is only 0.27 per cent. In view of the fact that an AGN is known to be present,
$3\sigma$ is regarded as an appropriate detection criterion. We further note
that the probability for a chance coincidence of a X-ray source with a radio 
source, based on geometrical considerations, is only $\sim 10^{-6}$. 
If no source is detected at the radio position, the 2$\sigma$ upper limit on 
the number of source photons is determined. The RASS exposure of the sources, 
used to calculate the corresponding count rates, is derived from the 
vignetting-corrected exposure map and is averaged over the area of the source.

The critical parameter in the calculation of a reliable upper limit is 
the local X-ray background at the source position. 
We calculate the local background by selecting a source-free box along 
the scanning direction of the telescope, but slightly offset from the 
supposed source position in the Survey. In this way, it is ensured that 
the background region has an exposure similar to that of the source. 

The unabsorbed fluxes are calculated from the count rates by assuming 
a simple power-law spectrum modified by Galactic absorption, which is 
parametrized by the neutral hydrogen column density ($N_{\rm H}$) towards 
the source (Stark et al. 1992) and the photoelectric absorption 
coefficients of Morrison \& McCammon (1983). The photon indices are 
chosen to be $\Gamma=1.9$ for galaxies and $\Gamma=2.15$ for quasars, 
which represent the class averages of the much larger samples of radio-loud 
X-ray sources of Brinkmann, Siebert \& Boller (1994) and Brinkmann et al.
(1995).  All luminosities are calculated in the rest frame of the sources, 
again using the class average power-law indices for the K-correction.

\subsection{Pointed ROSAT observations}

In addition to the data from the RASS, 34 sources were in the field of view 
of pointed {\it ROSAT} observations. A similar source detection procedure was 
applied to the data. However, due to the much smaller PSF of the PSPC in 
pointed observations, the standard extraction radius for source photons is 
reduced to 2.5 times the FHWM of the PSF for pointed mode at an energy of 
0.3 keV. The FWHM depends on the off-axis angle and ranges from $\sim$40 to 
$\sim$65 arcsec for on-axis and for 20 arcmin off-axis observations, 
respectively. For the sources that appeared extended due to their proximity 
(see below), the extraction radius was adjusted by hand to ensure the
inclusion of all photons from the AGN. For the sources in clusters of galaxies,
for which the AGN emission could not be spatially separated from the 
surrounding cluster emission (0255$+$05, 0625$-$53, 0915$-$11, 1246$-$41, 
1514$+$07), the standard extraction radius was used as well. This procedure is 
aimed at constraining the X-ray fluxes to the AGN emission as much as possible.
Fluxes and luminosities (see Table \ref{data}) are calculated assuming the 
same spectral parameters as for the RASS data. The errors in the fluxes are 
determined by photon statistics only and do not include uncertainties in the 
spectral parameters of the sources. Unless otherwise stated, we use the fluxes 
from the pointed observations in all further analysis.

\subsection{Comparison of the results from the RASS and from pointed 
observations}

\subsubsection{X-ray fluxes}

Five sources are detected in the pointings that had only upper flux limits 
from the RASS. Except for PKS 0325$+$02 (3C 88), the RASS upper limits are 
well above the flux determined from the pointed observations. For the radio 
galaxy 3C 88, however, the 95 per cent upper flux limit ($0.7\times 10^{-12}$
\ergs) is almost a factor of 2 lower than the actual flux derived from the 
pointed observation ($1.3\times 10^{-12}$ \ergs). Given this flux, the source 
would have been easily detectable in the RASS observation. Obviously, the 
soft X-ray flux of 3C 88 varied by at least a factor of two between the 
observations (i.e. within $\sim 12$ months). 

We compared the soft X-ray fluxes for the 29 sources for which data are 
available from both the RASS and a pointed observation and found agreement 
within the errors for most of them. The existing discrepancies can be 
explained by either variability or the influence of extended X-ray emission. 
For example, the fluxes of sources associated with clusters (indicated by `C' 
in Table 1) tend to be higher in the RASS than in the pointed observations. 
This is obviously due to the different extraction radii chosen for the two 
datasets.

Significant differences in the fluxes of the RASS and the corresponding pointed
observation also appeared for the quasars PKS 0403$-$13, 3C 279, PKS 1510$-$08
and both BL Lac objects. These discrepancies are most likely due to source 
variability. The source fluxes changed by factors of $\sim$2 (PKS 0403$-$13), 
$\sim$4 (3C 279) and $\sim$3.5 (PKS 1510$-$08) between the two observations. 
The flux of the BL Lac object PKS 0521$-$36 decreased by 50 per cent. The 
most extreme case is PKS 1514$-$24 (Ap Lib). It is clearly detected in the 
RASS with a count rate of $\sim 0.063$ count s$^{-1}$, but not in a 3-ks 
pointed observation. Based on the RASS count rate, $\sim 184$ photons are 
expected in the pointed observation, whereas the 95 per cent upper limit is 
only $\sim 9$ photons. This implies that the source flux decreased by at 
least a factor of 20 within 3 years. 

\subsubsection{X-ray extent}

Comparing the extent likelihoods of the sources in the RASS and the pointed 
observations, it turns out that the extent likelihood derived from the RASS 
observation is not a good predictor of the true source extent unless the value 
is very high. This is due to the complicated energy and angular dependence
of the PSF in the RASS, which can lead to a smearing of otherwise point-like 
sources, especially for bright objects. 11 sources had a likelihood for source 
extent greater than 10 in the RASS. Fortunately, 9 of them were also 
observed in pointed observations and, indeed, three objects turned out to be 
point-like in X-rays (3C 120, 3C 273, PKS 0620$-$52). In the case of 
PKS 0620$-$52 the RASS value obviously is affected by a second bright X-ray 
source only 3 arcmin away from the AGN. The remaining two sources, for which 
no pointed observation is available, either are associated with a known 
cluster (PKS 2104$-$25) or have a $B_{\rm gg}$ value (see Section 5.1.2) 
suggesting a dense environment (PKS 0442$-$28) and are thus likely to be 
extended in X-rays as well. 

\subsection{Previous X-ray observations}
 
There are two sources that were detected with the {\it Einstein} IPC,
but not in the RASS (0034$-$01, 2356$-$61). The derived upper limit fluxes 
from the Survey observation are well above the corresponding IPC fluxes. We 
converted the 0.5--4.5 keV IPC fluxes to the {\it ROSAT} energy band by 
assuming a photon index of 1.9.

For 3C 445, which has previously been claimed to be associated with the strong 
X-ray source 2A2220$-$022 (Marshall et al. 1978), we can derive only an upper 
limit on the flux from the Survey observation. Obviously, 2A2220$-$022 has 
been misidentified and it is most likely connected to the cluster A2440, 
which is located $\sim$ 50 arcmin north of 3C 445 and within the error box of 
2A2220$-$022. This has already been noted by Pounds (1990), who also reports a 
marginal detection of 3C 445 in an IPC observation with a flux comparable to 
our upper limit. Given the uncertainties in the spectral parameters of this 
source and in the flux conversion from the IPC to {\it ROSAT}, we use the 
{\it ROSAT} upper limit in our analysis. Similarly, 4U1716$-$01 is most 
likely not connected to 3C 353, as has been suggested by Forman et al. (1978) 
and Wood et al. (1984). Again, there is no X-ray source detected at the 
position of 3C 353 in the RASS, but extended emission is visible to the 
south-east of 3C 353, at a position which is consistent with the error box of 
4U1716$-$01 and which is probably due to a previously unidentified cluster of 
galaxies.

\section{Results}

\subsection{Soft X-ray properties}

The analysis of the sample resulted in 59 detections and 29 upper 
limits. Apart from PKS 1151$-$34, all quasars and BL Lac objects are detected, 
while for $\sim$40 per cent of the radio galaxies only upper limits to the 
X-ray flux could be determined. In Table \ref{data} we present the soft X-ray 
properties of the objects in our sample together with the relevant information 
from other wavebands.

In total 16 sources show significantly extended X-ray emission in either the
RASS or pointed observations. These sources are marked with `E' in Table 1.
Nine sources appear extended due to the X-ray emission from the associated 
cluster of galaxies (they are denoted with `E,C' in Table \ref{data}). 

The four closest sources in our sample (NGC 253, NGC 1068, Fornax~A, 
Centaurus~A) also show extended X-ray emission. NGC 253 is known to have a 
X-ray halo (Pietsch et al., in preparation), whereas an extended starburst 
component has been reported for NGC 1068 (Wilson et al. 1992; Ueno et al. 
1994). Finally, 3C 270 (Worrall \& Birkinshaw 1994) and Hercules A (Leahy 
1995) are extended in X-rays as well. The latter object will be discussed in 
more detail in Section 5.1.2.

\begin{figure}
\psfig{figure=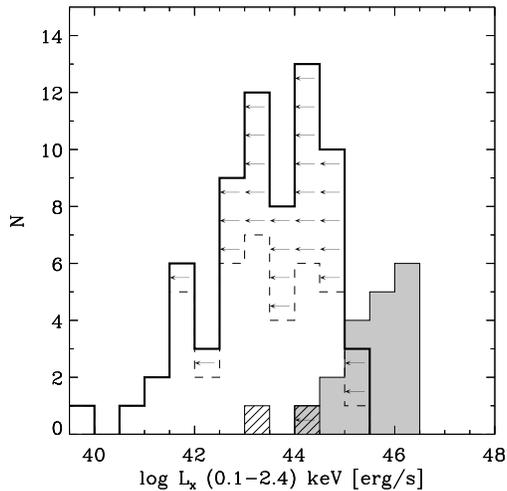,width=10cm}
\caption{\label{h_lx} Distribution of X-ray luminosities in the total {\it 
ROSAT}
energy band (0.1--2.4 keV) for quasars (dark), BL Lacs (hatched) and galaxies 
(solid line). Upper limits are indicated by arrows.}
\end{figure}

In Fig. \ref{h_lx} we show the soft X-ray luminosity distribution for 
the quasars, BL~Lac objects and galaxies in our sample. The galaxies and 
quasars are well separated with the quasars being brighter in X-rays. 
Moreover, the galaxies cover a wider range in X-ray luminosity than 
the quasars. While the quasars are confined to $10^{45}-10^{46}$ erg s$^{-1}$, 
the galaxy luminosities cover five orders of magnitude from 
$10^{40}$ erg s$^{-1}$ to $10^{45}$ erg s$^{-1}$, with most of the 
galaxies at the low-luminosity end being of the FR~I type.
 
\begin{figure}
\psfig{figure=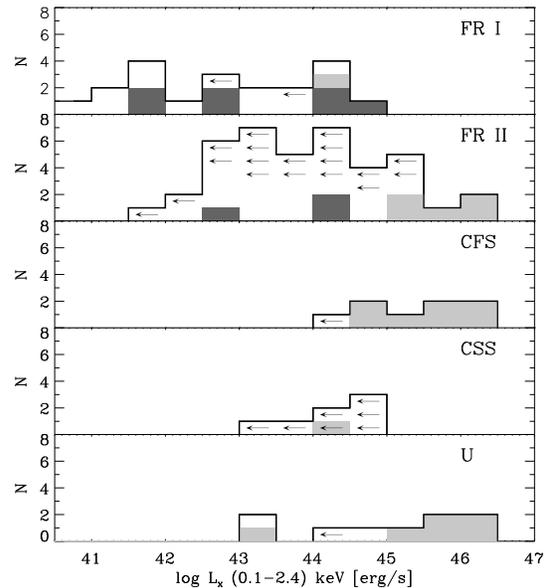,width=10cm}
\vspace{1cm}
\caption{\label{h_mo}
Distribution of the X-ray luminosities for the different radio morphological
subgroups. Quasars/BL Lacs and sources associated with 
clusters are represented by light and dark shading, respectively. Again upper 
limits are indicated by arrows. The morphological types are described in 
Table \ref{data}.}
\end{figure}

The distribution of the X-ray luminosities for the radio morphological classes 
is shown in Fig. \ref{h_mo}. At first sight, the FR~I type sources seem to 
be shifted towards lower X-ray luminosities compared with the FR~IIs. While 
there are some very X-ray-luminous FR~I galaxies that reach the top end of the
FR~II distribution, these are either associated with a cluster of galaxies or 
classified as a BL Lac object. The low-luminosity tail of the FR~I 
distribution overlaps with the region of normal elliptical and spiral galaxies
($L_{\rm x} = 10^{39}-10^{42}$ erg s$^{-1}$, Fabbiano 1989). On the other hand,
most of the highly X-ray-luminous FR~II type objects are also quasars. We 
applied {\sevensize ASURV} (LaValley, Isobe \& Feigelson 1992; Feigelson 
\& Nelson 1985; Isobe, Feigelson \& Nelson 1986) to test the hypothesis that 
the X-ray luminosity distributions for the FR~I and FR~II type objects are 
drawn from the same parent population using the narrow-line field radio 
galaxies only, i.e. excluding galaxies in clusters and broad-line objects 
(quasars, BLRGs). We find that the luminosity distributions of FR~I and FR~II 
type sources are now statistically indistinguishable. This is in contradiction 
to the results of Fabbiano et al. (1984), who concluded that FR~IIs are on 
average brighter in X-rays than FR~Is, although it should be noted that
they included broad-line galaxies in their analysis.

The fraction of upper limits among the FR~IIs is much higher than among the 
FR~Is. To first order, this is due to a distance effect, because the FR~I 
galaxies are on average at a smaller redshift in this radio flux limited 
sample, since only the most radio luminous galaxies, i.e. FR~IIs, can be seen 
at higher redshifts. But even considering only those objects that cover the 
same redshift range ($0.03 \le z \le 0.112$, which excludes all low-luminosity
FR~Is), the detection rate of FR~I field radio galaxies (using the RASS 
results) still tends to be higher (FR~I: 4/6 = $67(\pm33)$ per cent; FR~II: 
4/15 = $27(\pm13)$ per cent). Interestingly, the luminosity distributions for 
the restricted subsamples are now different at the 95 per cent confidence 
level: compared with the FR~IIs, the X-ray luminosites of FR~Is are {\it 
higher} on average.

The compact flat-spectrum (CFS) and the unclassified (U) sources show the 
highest X-ray luminosities. In addition, these morphological classes also 
show very high X-ray detection rates. Most of the CFS and many U sources are 
quasars or BL Lac objects, which were just resolved in the radio observations 
(Morganti \eta 1993), but impossible to classify in terms of radio morphology. 
Finally, the lack of detections of compact sources with a steep radio spectral
index (either compact steep-spectrum (CSS) or GHz-peaked (GPS) sources), 
already pointed out from the analysis of larger samples (Brinkmann et al. 
1994, 1995), is likely to be due to the high redshifts at which these sources 
are usually found. In fact, there are only two out of 18 radio galaxies 
detected in the redshift range of the CSS sources (typically {\it z}$>$ 0.2), 
and both show broad lines. Nevertheless, it is worth mentioning that the only 
quasar not detected in soft X-rays is classified as a CSS source as well.

\subsection{Undetected sources}

\begin{figure}
\psfig{figure=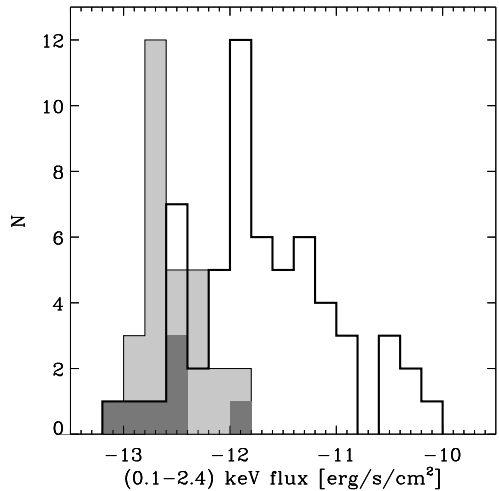,width=10cm}
\caption{\label{h_fx} X-ray fluxes of all 88 2-Jy sources in the total 
0.1--2.4 keV energy band. Detected sources are plotted with a thick 
line while sources with upper limits are represented by the light shaded area.
Sources that are detected in pointed observations only are indicated by the 
dark shaded area.}
\end{figure}

In Fig. \ref{h_fx} we plot the flux distribution of all sources in the 0.1--2.4
keV energy band. Apart from the sources that were detected in pointed 
observations with longer exposure (dark shaded), detections and upper limits 
are well separated. Nevertheless, the two distributions show an 
overlap which reflects the inhomogeneous exposure of the sources in 
the RASS and the variations in the Galactic $N_{\rm H}$ value. 
The sources detected in pointed observations only are included in the analysis 
with their RASS upper limit fluxes in this section, because of the generally 
much longer exposure in pointings compared with the RASS observations.

One possible reason for a non-detection in soft X-rays could be an 
unusually high value of Galactic absorption along the line of sight. 
A Kolmogorov-Smirnov test gives a probability of 80 per cent that the 
$N_{\rm H}$ distributions of detected and undetected sources were drawn from 
the same parent population. Thus there is no significant bias towards high 
$N_{\rm H}$ values among the non-detected sources. 

The varying exposure of the sources during the RASS does not explain 
why some sources are detected while others are not. With very few 
exceptions, the distributions of the exposures for detected and non-detected 
sources are very similar. 

\begin{figure}
\psfig{figure=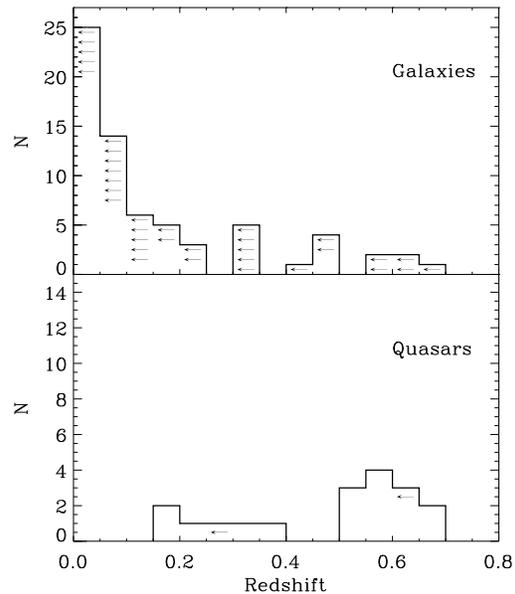,width=10cm}
\vspace{1cm}
\caption{\label{h_z}
Redshift distributions for galaxies and quasars. Sources that only have an 
upper limit from the RASS observation are indicated by arrows.}
\end{figure}

The last possibility, apart from intrinsic source properties, is a 
distance effect. In Fig. \ref{h_z} we plot the redshift distributions for 
galaxies and quasars with upper limits indicated by arrows. In the case of the 
galaxies it is obvious that the detections and the upper limit sources have 
markedly different redshift distributions. The majority of the detected 
sources have redshifts smaller than 0.1 and there are only two galaxies 
detected beyond {\it z} = 0.25 (1602$+$01, 1938$-$15). 

\subsection{Correlations}

In the following sections we compare the X-ray luminosities of the 
objects with their emission properties in the radio and the optical 
wavebands. The correlation and regression analysis including 
the upper limits was done with {\sevensize ASURV} (LaValley et al. 1992). 

\subsubsection{X-ray -- Radio correlations}

\begin{figure}
\psfig{figure=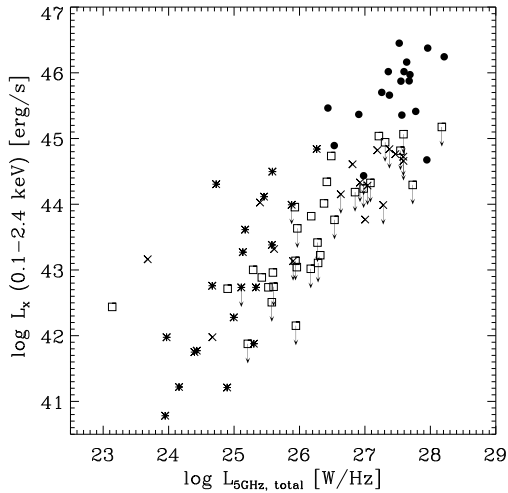,width=10cm}
\caption{\label{lx_lt} Integrated 0.1-2.4 keV  X-ray luminosity versus total
5-GHz radio luminosity ($\bullet$=quasars, $\ast$=FR~I galaxies, $\Box$= FR~II 
galaxies, $\times$=other galaxies, arrows indicate upper limits).}
\end{figure}

\begin{figure}
\psfig{figure=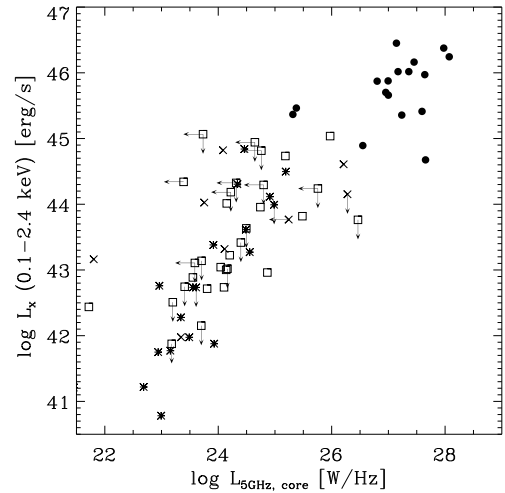,width=10cm}
\caption{\label{lx_lc} X-ray luminosity versus 5-GHz core luminosity 
($\bullet$=quasars, $\ast$=FR~I galaxies, $\Box$= FR~II galaxies, 
$\times$=other galaxies, arrows indicate upper limits).}
\end{figure}

Figs \ref{lx_lt} and \ref{lx_lc} show the plots of the 0.1--2.4 keV 
X-ray luminosity \lx versus the total 5-GHz luminosity \lt and the 5-GHz core 
luminosity \lc, respectively. The results of the correlation and regression 
analysis for various object classes are summarized in Table \ref{reg}, where 
we also give the respective probabilities that the observed correlations 
arise by chance. In general, the derived parameters have large uncertainties 
(all errors are 1$\sigma$ in Table \ref{reg}), which is expected given the 
low number of objects and the large fraction of upper limits. Nevertheless, 
an inspection of the figures clearly suggests the presence of a correlation 
between \lx and both \lt and \lc for quasars as well as for radio galaxies. 
This is confirmed by a simple correlation analysis, although, statistically, 
the correlation probability for quasars is only marginally significant. This 
is partly due to the low number of objects. Note that the significance of the 
\lx -- \lc correlation is higher than that of the \lx -- \lt correlation for 
both object classes.

In interpreting these correlations one has to consider selection effects, 
in particular the dependence of the total radio luminosity on redshift 
{\it z}, artificially introduced by the radio flux limit of the original 
Wall \& Peacock sample. \lx and \lc do not depend a priori on redshift, 
because of the inclusion of upper limit values. Further, the correlations of 
\lx with \lt and \lc are not mutually independent since \lc is also 
correlated with \lt. Usually, a partial correlation technique is applied to 
analyse a many-variable problem like this and to disentangle the real 
correlations from those introduced by the individual dependences amongst 
the variables. However, these methods fail to account properly for censored 
data. To deal with these, a new procedure, based on Kendall's $\tau$, was 
developed to calculate the partial correlation coefficient and to estimate 
its significance in the presence of upper limits (Akritas \& Siebert 1996). 

We applied this new procedure to estimate the influence of the \lt -- \lc and 
\lt -- {\it z} relations on the correlation between \lx and both \lt and \lc.  
In the case of the quasars, the \lx -- \lt correlation seems to be strongly 
affected by both the redshift bias and the \lc -- \lt correlation. It turns 
out that the correlation is no longer statistically significant once both 
selection effects are properly accounted for. The \lx -- \lc correlation is 
much less affected and the probability of erroneously rejecting the null 
hypothesis of no correlation is $\la 4$ per cent, which is acceptable given 
the low number of objects. We thus conclude that there is indeed a correlation 
between \lx and \lc for quasars and that the \lx -- \lt correlation is most 
likely an artefact of the redshift bias and/or the strong relation between 
\lt and \lc in our sample. The remarkably flat regression slope for quasars 
in the \lx -- \lc correlation is consistent with previous findings. Kembhavi 
\eta (1986) give $0.465\pm0.04$ as best-fitting slope to a sample of 86 
quasars with measured radio core fluxes including X-ray upper limits, and 
Baker, Hunstead \& Brinkmann (1995) report $0.36\pm0.1$ for a large sample of 
steep-spectrum quasars. However, these results are inconsistent with the 
findings of Browne \& Murphy (1987), who derive a regression slope of 
0.75 for a sample of 135 quasars. The difference may be explained by 
the heterogeneity of their sample and the selection effect they 
introduce by not including upper limits in their analysis. 

The results for the radio galaxies are similar. The \lx versus \lc correlation 
remains highly significant in the partial correlation analysis, whereas 
the probability of no correlation between \lx and \lt increases from 0.03 
per cent to 7.5 per cent once the effect of redshift is taken into account. 
Given the much larger sample size compared with the quasars, this result 
provides evidence for the \lx -- \lt correlation being introduced by the 
redshift bias. Interestingly, the regression parameters for the \lx -- \lt 
and the \lx -- \lc correlations are almost identical ($a \sim 0.6\pm 0.1$). 
Fabbiano \eta (1984) and Brinkmann \eta (1995) find a slightly steeper 
regression slope in the case of the galaxies ($\sim0.8\pm 0.2$), but 
consistent within the errors.

We repeated the correlation and regression analysis for the two FR classes 
separately. Unfortunately, the large number of upper limits among the FR~II 
radio galaxies does not allow a reliable determination of the regression 
parameters, and the \lx -- \lt correlation turns out to be insignificant in the
partial correlation analysis. Note that the \lx -- \lc correlation 
remains significant. Formally, we find strong correlations between \lx and 
both \lc and \lt for the FR~I galaxies. The \lx --\lt correlation, however, 
is only marginally significant in the partial correlation analysis, whereas 
the \lx -- \lc correlation is obviously not influenced by any of the 
above-mentioned selection effects, and the regression slope is close to unity. 
Both findings indicate a contribution to the X-ray emission from the active 
nucleus in FR~I galaxies. 

\begin{table*}
\begin{minipage}{15.5cm}
\caption{\label{reg}Results of the correlation and regression analysis.}
\begin{tabular}{@{}cccrrrccccc}
 $Y$ & $X$ & Class & N & UL$_{\rm Y}$ & UL$_{\rm X}$ & {\it a} & {\it b} & P & 
P$_{\rm L_c/L_T}$ & P$_{\rm z}$\\
 &  &  &     &  &  &  &  \\
$\log L_{\rm x}$ & $\log L_{\rm total}$ & QSO  & 18 & 1 & 0 &  0.59$\pm$0.26 & 
29.54$\pm$7.25 & 0.081 & 0.749 & 0.271 \\
        &                  & GAL &  68 & 28 & 0 & 0.58$\pm$0.10 & 
28.14$\pm$2.67 & 0.0003 & 0.004 & 0.075 \\
        &                  & FR~I  & 20 & 2 & 0  &  1.37$\pm$0.29 & 
8.52$\pm$7.28 & 0.0013 & 0.067 & 0.082 \\
        &                  & FR~II  & 31 & 17 & 0  &  0.45$\pm$0.20 & 
31.30$\pm$5.11 & 0.0053 & 0.014 & 0.32 \\
        &                  & ALL  & 88 & 29 & 0 & 0.88$\pm$0.08 & 
20.50$\pm$2.19 & $\ll 10^{-6}$ \\
 &  &  &   &  &   \\
$\log L_{\rm x}$ & $\log L_{\rm core}$  & QSO  & 17 & 0 & 0 & 0.30$\pm$0.14 & 
37.65$\pm$3.79 &  0.017 & 0.039 & 0.017\\
        &                 & GAL &  59 & 20 & 10 & 0.63$\pm$0.12 & 
27.66$\pm$2.85 & $<10^{-6}$ & $<10^{-4}$ & $<10^{-4}$ \\
        &                  & FR~I  & 20 & 2 & 0 &  1.00$\pm$0.18 & 
19.08$\pm$4.40 & 0.0002 & 0.007 & 0.014\\
        &                  & FR~II  & 30 & 16 & 8 &  0.58$\pm$0.26 & 
28.94$\pm$6.28 & 0.0011 & 0.012 & 0.016\\
        &                  & ALL  & 78 & 20 & 10 &  0.77$\pm$0.06 & 
24.40$\pm$1.48 & $\ll 10^{-6}$ \\
 &   &  &  &  &   \\
$\log L_{\rm x}$ & $\log L_{\rm opt}$  & QSO  & 18 & 1 & 0 &  0.84$\pm$0.16 & 
19.54$\pm$4.87 & 0.0006 & & 0.0017\\
 &   &  &  &  &   \\
$\log L_{\rm x}$ & $\log L_{\rm [O~III]}$ & QSO  & 18 & 1 & 0 &  0.92$\pm$0.20 
& 5.65$\pm$8.58 & 0.0021 & & 0.007\\
      &                     & FR~II   & 30 & 16 & 0 &  0.41$\pm$0.12 & 
25.97$\pm$4.93 & 0.19 \\
      &                     & QSO \& FR~II & 48 & 17 & 0 &  0.92$\pm$0.11 & 
4.86$\pm$4.54 & $\ll 10^{-6}$ & & 0.005\\
      &                     & FR~I   & 19 & 2 & 12  &  0.67$\pm$0.30 & 
16.61$\pm$11.78 & 0.025 & & 0.069\\

\end{tabular}

\medskip

$X$ and $Y$ denote the independent and the dependent variables, respectively. 
N is the total number of objects while UL$_{\rm X}$ and UL$_{\rm Y}$ are 
the number of upper limits in the independent and the dependent variables. 
Linear regression has been performed assuming $\log Y = a \log X + b$. All 
errors are 1$\sigma$. P is the probability that no correlation is present, 
whereas P$_{\rm L_c/L_T}$ and P$_{\rm z}$ denote the same probability but 
with the effects of the \lc -- \lt correlation and redshift excluded, 
respectively.  

\end{minipage}
\end{table*}

\subsubsection{X-ray -- optical correlations}

\begin{figure}
\psfig{figure=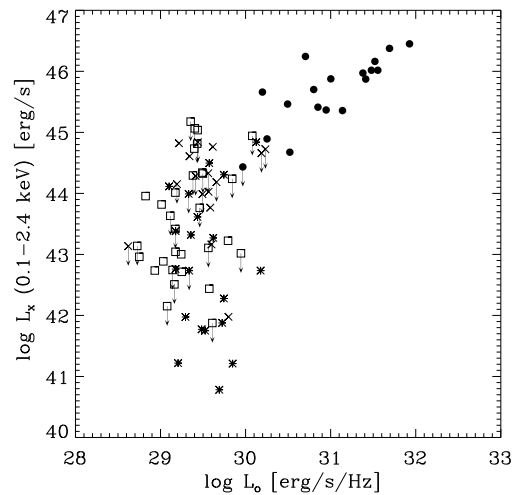,width=10cm}
\caption{\label{lx_lo} Optical luminosity versus soft X-ray luminosity 
($\bullet$=quasars, $\ast$=FR~I galaxies, $\Box$= FR~II galaxies,
$\times$=other galaxies, arrows indicate upper limits).}
\end{figure}

In Fig. \ref{lx_lo} we plot the soft X-ray luminosity versus the optical 
luminosity. The rest-frame optical luminosity was calculated from the 
$V$ magnitudes, using the formula of Allen (1976) for flux conversion. 
We applied an energy power-law index of $\alpha=0.5$ for the K-correction. 
There is an excellent correlation between the optical and the X-ray 
luminosities for quasars, which is not affected by the redshift bias. The 
best-fitting regression slope is $0.84\pm0.16$. The correlation found is 
slightly steeper than the value obtained from the much larger 
{\it ROSAT}/Condon quasar sample (Brinkmann et al. 1995, $0.75\pm0.12$), but 
is still within the mutual errors. Boyle \eta (1993) derive a slope of 
$0.88\pm0.08$ from an analysis of the optical and the soft X-ray luminosity 
functions of an X-ray-selected sample of quasars. Avni \& Tananbaum (1986) 
give 0.8 as the best-fitting regression slope for a sample of 94 
{\it Einstein}-detected quasars and 60 upper limits, whereas Wilkes \eta 
(1994) find a slightly flatter slope (0.71) for an even larger sample, but 
still in agreement with our results. No correlation between the optical and 
the X-ray luminosities is observed for the radio galaxies.

\begin{figure}
\psfig{figure=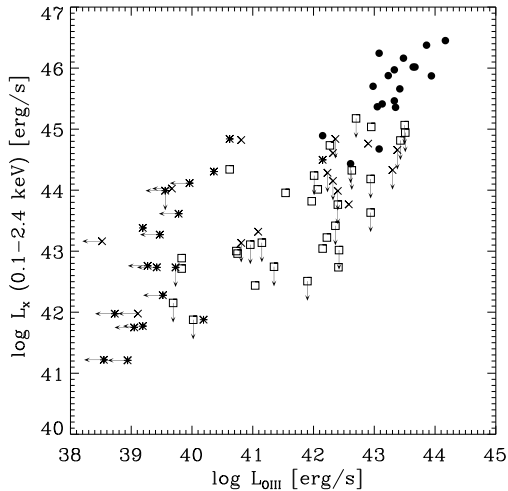,width=10cm}
\caption{\label{lx_o3} [O\,{\sevensize III}] line luminosity versus X-ray luminosity 
($\ast$= FR~I, $\Box$= FR~II, $\times$=other galaxies, $\bullet$= quasars, 
arrows indicate upper limits).}
\end{figure}

In Fig. \ref{lx_o3} we plot the soft X-ray luminosity versus the 
[O\,{\sevensize III}]$\lambda5007$ line luminosity. There is a highly significant 
correlation visible for the quasars and the regression slope is consistent 
with unity. Partial correlation analysis shows that this result is not 
spurious due to the possible correlations of \lx and 
$L_{\rm O\,\sevensize III}$ with \lt, \lc or $z$. Therefore \lx seems to be 
physically related to $L_{\rm O\,\sevensize III}$, which argues for the 
narrow-line gas being photionized by the AGN. Although we fail to find a 
significant correlation for the FR~II radio galaxies as well, we note a 
smooth transistion from quasars to FR~IIs, whereas FR~I galaxies definitely 
show different behaviour with, in general, much lower [O\,{\sevensize III}] 
luminosity compared with their X-ray emission.

\section{Discussion}

Two main points are apparent from the above analysis. First, there is 
a very large range in X-ray luminosity over which the radio galaxies can 
be observed. This is probably the effect of different mechanisms at 
work in producing the X-rays. Secondly, anisotropic X-ray emission from the 
active nucleus is likely to be present also in radio galaxies as indicated 
by the correlation of X-ray luminosity with radio core power in FR~I galaxies. 
In the next sections we discuss the role of the emission from the active 
nucleus in different object classes, the implications for unified schemes and 
possible mechanisms responsible for the X-ray emission in radio galaxies.

\subsection{Sources of X-ray emission from radio galaxies}

\subsubsection{The contribution of the AGN}

The correlation of \lx with \lc for both quasars and radio galaxies suggests 
that the X-ray emission is related to the radio emission from the core of the 
AGN, for example by synchrotron self-Compton or inverse Compton processes in 
the radio jet. The flat slope of the \lx -- \lc correlation for quasars might 
be due to unresolved radio emission in the cores of the quasars, which is not 
related to the X-ray emission. This is plausible given the high redshift
of the objects and has already been pointed out by Kembhavi et al. (1986).
Indeed, they find a steepening of the regression slope as they pass from 
quasars with resolved cores to unresolved radio sources with flat or inverted 
spectra, which presumably are dominated by `real' core emission. 
   
\begin{figure}
\psfig{figure=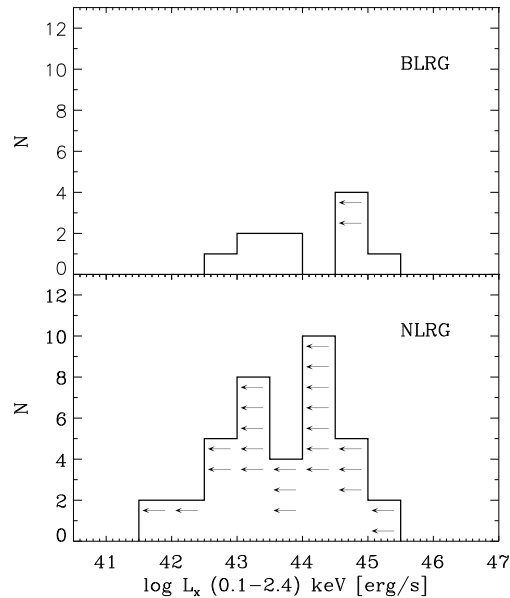,width=10cm}
\vspace{1cm}
\caption{\label{h_bl} X-ray luminosity distribution for broad-line and 
narrow- or weak-lined radio galaxies.}
\end{figure}

A subset of radio galaxies in our sample shows broad emission lines (Tadhunter 
et al. 1993; Shaw et al. 1995). In Fig. \ref{h_bl} we plot the distribution 
of X-ray luminosities for galaxies classified as broad-line (BLRG, upper panel)
and the ones with narrow or weak lines (lower panel). We excluded the FR~I 
galaxies because none of the BLRGs is unambiguously classified as FR~I. The 
histogram shows the higher detection rate and, on average, higher X-ray 
luminosity of the BLRGs compared with the other radio galaxies. A statistical 
test with {\sevensize ASURV} reveals that the two distributions are 
different at the 99.5 per cent confidence level.

This result is consistent with the unified scheme for high-power radio sources
(Barthel 1989). In this scheme, it is proposed that we see the nucleus 
directly in the case of quasars, whereas in the narrow-line radio galaxies
(NLRGs) the nucleus is `hidden' by a combination of relativistic beaming 
effects and dust/gas obscuration. BLRGs are supposed to be observed at 
intermediate angles. Since the obscuration is likely to be significant at 
soft X-ray energies, we would expect a higher X-ray detection rate amongst 
the quasars and BLRGs, because they are less obscured. Further evidence that 
the BLRGs are intermediate objects can be derived from the fact that the 
optical spectra of BLRGs do show differences compared with those of quasars, 
including redder continua and steeper Balmer decrements (e.g., Osterbrock, 
Koski \& Phillips 1976). This can be interpreted in terms of partial 
obscuration. Finally, Turner \& Pounds (1989) conclude from an {\it EXOSAT} 
study of emission-line AGNs that intrinsic absorption is quite common in 
BLRGs, which again argues for BLRGs being intermediate objects. On the other 
hand, although most of our BLRGs do show red continua, some of them may also 
be nearby quasars where the host galaxy can be seen. This would also explain 
the observed deficiency of nearby quasars in our sample: there are 24 FR~II 
radio galaxies with a redshift $z \le$ 0.2 in our sample. Assuming that all 
FR~IIs are `misdirected' quasars and, further, that FR~IIs with angles to the
line of sight below $45\degr$ are classified as quasars, we would expect to 
see 7$\pm$2.7 quasars with $z \le$ 0.2, whereas 3 are observed. The 
discrepancy, although marginal, increases if one takes into account that our 
sample may be biased towards core-dominated sources due to the relatively 
high radio selection frequency.

\begin{figure}
\psfig{figure=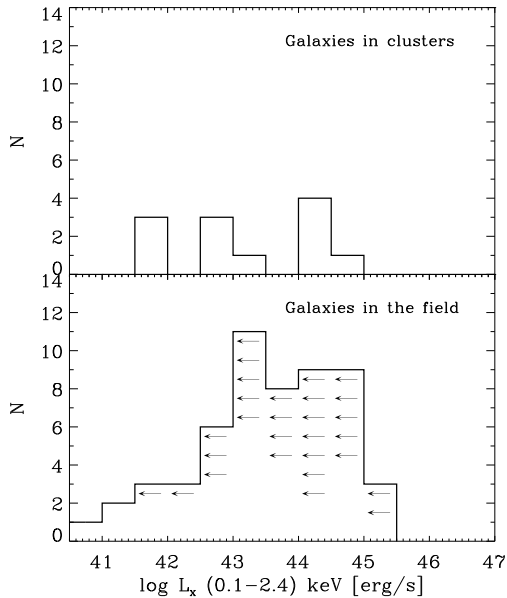,width=10cm}
\vspace{1cm}
\caption{\label{h_cl} X-ray luminosity distributions for radio galaxies in 
the field and those associated with a cluster.}
\end{figure}

Although evidence for anisotropic nuclear X-ray radiation in radio galaxies is 
mainly restricted to the broad-line sources, such radiation could contribute 
to the narrow- or weak-lined radio galaxies as well. Monte Carlo simulations 
(Morganti \eta 1995) have shown that the observed differences in the radio 
core detection rates of FR~I and FR~II radio galaxies (Morganti \eta 1993) 
can be explained if the jets of FR~I galaxies are characterized by on average 
lower Lorentz factors compared with FR~II galaxies. If there is also an 
anisotropic contribution to the X-ray emission in these galaxies, which is 
affected by the same Lorentz factors as the radio emission, we expect a 
higher X-ray detection rate for the FR~I galaxies, since the radiation is 
then emitted in a broader beam and is thus more likely to be observed. 

In Section 4.1 we have presented indications for a higher detection rate
and higher X-ray luminosities of FR~I type galaxies compared with FR~IIs, in 
carefully chosen subsamples to avoid various selection effects. Unfortunately,
the significance of the results is limited by the small number of objects 
and, although we restricted the analysis to a small range in redshift, there 
is still a tendency for FR~II galaxies to be at higher redshifts. Nevertheless,
the model of an anisotropic and less beamed X-ray component in 
FR~I galaxies compared with FR~II represents a viable explanation of our
data, but the statistical evidence is not particularly strong. Larger samples 
of FR~I and FR~II galaxies are required to decide if the soft X-ray 
detection rate and luminosities of FR~I galaxies are really higher.

We note, however, that evidence for non-thermal X-ray emission from
low-power radio galaxies has recently been reported by Worrall \& Birkinshaw
(1994). After deconvolving the extended and unresolved X-ray emission 
components, they find a correlation with slope unity between the unresolved
X-ray luminosity and the 5-GHz core luminosity. This can be readily explained
if the X-rays originate from the inner regions of a parsec-scale radio jet
(Worrall \& Birkinshaw 1994).

\subsubsection{ISM}

Thermal bremsstrahlung from a hot interstellar medium is known to be an 
important component of the X-ray emission from galaxies (e.g. Canizares 1987).
This emission may come from diffuse gas distributed on very different spatial 
scales: clusters, small groups or galactic haloes. With {\it ROSAT} it has 
become feasible to prove the presence of extended thermal emission even in
isolated and/or distant radio galaxies. Recently, evidence for an extended 
emission component has been reported for several distant radio galaxies 
(e.g. Worrall et al. 1994; Crawford \& Fabian 1995). However, apart from the 
sources already mentioned in Section 4.1, no evidence for extended emission 
has been found in our sample, but, as already pointed out in Section 3.3.2, 
the information on the X-ray size of the sources is not conclusive in the RASS.

Various models have been proposed to explain the correlations of \lx with 
\lt and \lc for radio galaxies in the context of a thermal origin of the 
X-ray emission. For example, Feigelson \& Berg (1983) relate the correlation 
of \lt with \lx in a sample of powerful (mostly FR~II) 3CR galaxies to the 
presence of hot X-ray emitting gas in the putative cluster environment of 
these objects, which confines the radio lobes and thus also increases the 
radio emission due to inhibited adiabatic expansion losses. Fabbiano, Gioia 
\& Trinchieri (1989) suggested that the hot ISM in radio galaxies fuels the 
AGN via accreting cooling flows and thus the strength of the radio core 
emission is related to the amount of gas (and hence X-ray emission) in those 
objects. The strong correlations found for FR~I radio galaxies in our sample 
could support these arguments, since nearby FR~Is are frequently found in 
cluster environments (e.g. Prestage \& Peacock 1988). We note, however, that 
the correlations also hold for the apparently isolated nearby FR~I objects 
and that the \lx -- \lt correlation may be an artefact of the redshift bias 
(see Section 4.3.1) 

In order to investigate further the possible contribution of thermal emission 
from a hot intracluster medium, we plot in Fig. \ref{h_cl} the distribution 
of the X-ray luminosities for the radio galaxies in clusters (upper 
panel) and for field galaxies (lower panel). The X-ray detection rate of the 
cluster sources is much higher than that of the field glaxies, which is most 
likely explained by the on average lower redshifts of the cluster sources 
compared with the field galaxies. We derive a (marginally significant) 
$\sim 85$ per cent probability that the luminosity distributions of the two 
subsamples are different, with the cluster sources showing slightly higher 
X-ray luminosity. In order to quantify better the contribution of the cluster
medium to the X-ray emission, we have collected the values of the so-called
$B_{\rm gg}$ parameter, which is a measure of the distribution of galaxies 
around a given source (Lilly \& Prestage 1987; Prestage \& Peacock 1988; 
Yates, Miller \& Peacock 1989). There does not appear to be any correlation 
between the $B_{\rm gg}$ parameter and the X-ray luminosity. However, a 
possible positive correlation may be masked by the fact that the source 
detection algorithm systematically misses parts of the extended X-ray flux 
for sources with high values of $B_{\rm gg}$. We therefore conclude that the 
cluster X-ray emission, despite our efforts to isolate the AGN, still may 
contribute to the X-ray flux in some objects, but that the cluster emission 
is on average not the {\it dominant} source of X-ray emission in our sample 
of radio galaxies.

A group of radio galaxies in the sample showing high X-ray luminosities
($L_{\rm x} \ga 10^{43}$ erg s$^{-1}$) appear to be isolated objects without 
strong emission lines in the optical spectrum. The most extreme case is 
certainly Hercules A.\footnote{Whether or not Her A is associated with a 
cluster of galaxies has not yet been unambiguously decided. Contradictory 
results, based on galaxy counts or spatial correlation functions, have been 
reported in the literature (e.g. Yates et al. 1989; Allington-Smith et al. 
1993).} Other similar objects are 0305$+$03, 0325$+$02, 1637$-$77 and 
1954$-$55.
 
Ponman et al. (1994) have recently reported the discovery of a so-called 
fossil group, a single elliptical galaxy that is considered to be the result 
of the merging process of a compact group. This merging is believed not to 
affect the X-ray halo of the group (Ponman \& Bertram 1993) and the 
ellipticals formed in this way will have a high X-ray luminosity though they 
appear isolated. The typical X-ray luminosities of these groups are 
$L_{\rm x} = 10^{41} - 4\times 10^{43}$ erg s$^{-1}$ (Ponman \& Bertram 1993; 
Ebeling, Voges \& B\"ohringer 1994). These luminosities are low compared with 
the X-ray luminosity observed for Her~A, both by {\it ROSAT} ($L_{\rm x} = 
6\times 10^{44}$ erg s$^{-1}$) and by {\it Einstein} ($L_{\rm 0.5-4.5 keV} = 
1.4\times 10^{44}$ erg s$^{-1}$, Feigelson \& Berg 1983). However, good 
agreement between Her~A and the objects studied by Ponman et al. (1994) seems 
to be present with respect to some other source characteristics, like the 
extension of the X-ray emission and the effective radius ($R_{\rm e}$) of 
the optical galaxy. Furthermore, evidence for a recent or ongoing merger has 
been reported from optical observations (Sadun \& Hayes 1993). Perhaps 
additional X-ray emission associated with the active nucleus has to be 
considered in the case of Her~A, given the high radio power of this galaxy. 

The process suggested by Ponman et al. (1994) can be important in 
understanding the X-ray emission from isolated objects although 
spatial information is necessary to test the hypothesis for other 
radio sources. 

\begin{figure}
\psfig{figure=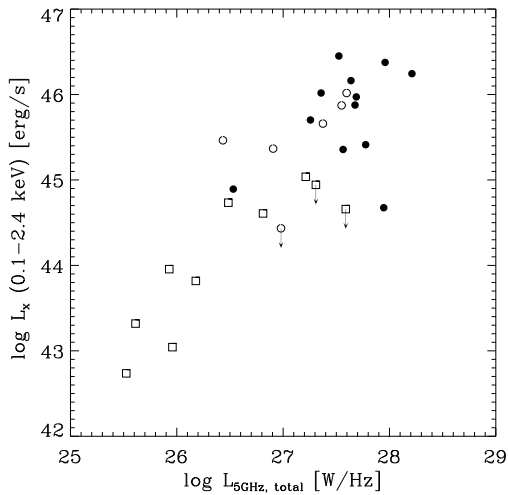,width=10cm}
\caption{\label{lx_lt_bl}
Soft X-ray luminosity versus radio luminosity for BLRGs (squares) and
quasars (circles). Flat -spectrum objects with $\alpha^{\rm 4.8}_{\rm 2.7} 
\le 0.5$ are plotted with filled symbols.} 
\end{figure}

Finally, the low-power tail of the distribution of the FR~I galaxies
($L_{\rm x} = 10^{41} - 10^{42}$ erg s$^{-1}$, see Fig. \ref{h_mo}) is 
consistent with the X-ray emission from normal early-type galaxies as found 
by Forman, Jones \& Tucker (1985), and in fact there are two objects, Cen A 
and For A, that are included in both samples. Forman et al. (1985) find 
typical X-ray luminosities ranging from $L_{\rm x} \sim 10^{39}$ to $\sim 
8\times 10^{41}$ erg s$^{-1}$ and a frequent occurrence of hot coronae. The 
dominant X-ray emission in these cases has been interpreted as thermal 
emission from hot gas which has been accumulated as a result of mass loss 
from evolved stars. 

\subsection {X-rays and the nature of the radio-optical correlations}

It has been known for some time that the optical emission-line luminosity 
is closely correlated with the radio power for radio galaxies (e.g. Hine 
\& Longair 1979; Rawlings \& Saunders 1991). It is generally assumed that 
this correlation arises because the strength of the EUV/X-ray continuum 
responsible for photoionizing the warm gas clouds increases with radio power. 
If this AGN photoionization model holds, we would expect to observe a general 
increase in the strength of the AGN X-ray emission with radio power.

The problem we face in testing the X-ray photoionization model is that part 
of the X-ray emission from the galaxies in our sample is likely to be due to 
thermal ISM emission (see above). We therefore concentrate on the broad-line 
objects (BLRGs and quasars) for which we are confident that most of the X-ray 
emission is emitted by the AGN. Fig. \ref{lx_lt_bl} shows a plot of X-ray 
luminosity against radio power for the combined sample of BLRGs and quasars. 
There is a general increase in the X-ray luminosity moving from BLRGs
to quasars, and the level of this increase is just as expected
on the basis of the photoionization models and the radio-optical correlations.
Indeed, there is almost a direct proportionality between the X-ray
and the emission-line luminosities (see Fig. \ref{lx_o3} and Table \ref{reg}). 
The results are generally consistent with the idea that the nuclei of BLRGs 
are the illuminating AGNs in the lower power/lower redshift objects, while 
quasars are the illuminating AGNs in the higher power/higher redshift objects.

One possible complication is that the flat-spectrum objects in our sample
may be dominated by the relativistically beamed core component, which may 
not have a large influence on the ionization of the gas because of the 
narrowness of the beam. However, we see from Fig. \ref{lx_lt_bl} that the 
flat- and steep-spectrum sources follow approximately the same distributions 
in the diagram, so the beamed component does not appear to affect the result. 

\section {Conclusions}

We have presented an analysis of the soft X-ray properties of a complete 
sample of 88 radio sources (68 galaxies, 18 quasars and 2 BL Lac objects)
derived from the Wall \& Peacock (1985) sample. The X-ray data, taken from 
the {\it ROSAT} All-Sky Survey, the {\it ROSAT} public data archive, and 
previous {\it Einstein} measurements, finally resulted in 59 detections and 29 
upper limits. While all but one quasar is detected, $\sim 40$ per cent of the 
galaxies only have upper limits to the X-ray flux from the Survey. The 
detection rate of the galaxies is clearly redshift dependent.

We find strong correlations between \lx and \lc for all object classes, 
which can be interpreted in terms of a significant contribution of the 
AGN to the X-ray luminosity. A partial correlation analysis shows that the 
correlation of \lx with \lt is probably an artefact of the redshift dependence 
of both luminosities and/or the strong correlation of \lt with \lc. 
Furthermore, there are correlations present between \lx and the optical 
continuum as well as with [O\,{\sevensize III}]-line luminosity for quasars.

The high detection rate of BLRGs and quasars, as well as the observed
correlation between X-ray and both the radio core and the 
[O\,{\sevensize III}] luminosities suggests that the X-ray emission of 
these objects is dominated by the contribution of the AGN. This is consistent 
with orientation-dependent unification schemes for powerful radio sources. 
Excluding the BLRGs and the quasars, the case for an AGN component of the 
X-ray emission is less clear, although the strong correlation between \lx and 
\lc in FR~I radio galaxies argues for a nuclear contribution to the X-ray 
emission. 

The X-ray emission of the narrow- or weak-lined radio galaxies is more likely 
dominated by the contribution of the host galaxies, groups or clusters. 
The X-ray luminosity of some FR~I galaxies is consistent with the emission 
from the ISM of normal ellipticals, whereas for the high-luminosity objects 
additional mechanisms are required. There are some isolated galaxies with a
relatively high X-ray luminosity ($L_{\rm x} > 10^{43}$ erg s$^{-1}$), which
show no strong optical emission lines. These objects could be single 
elliptical galaxies resulting from the merging process of a compact group, 
as suggested by Ponman et al. (1994). 

\section*{Acknowledgments}
The {\it ROSAT} project is supported by the Bundesministerium f\"ur Bildung, 
Wissenschaft, Forschung and Technologie (BMBF) and by the Max-Planck Society. 
JS and WB thank their colleagues from the {\it ROSAT} group for their support. 
This research has made use of the NASA/IPAC Extragalactic Data Base (NED) 
which is operated by the Jet Propulsion Laboratory, California Institute 
of Technology, under contract with the National Aeronautics and Space 
Administration.

\end{document}